# Polariton Topological Transition Effects on Radiative Heat Transfer

Cheng-Long Zhou,[1,2] Xiao-Hu Wu,[3] Yong Zhang,[1,2] Hong-Liang Yi,[1,2,*] and Mauro Antezza[4,5,†]

[1]*School of Energy Science and Engineering, Harbin Institute of Technology, Harbin 150001, P. R. China*
[2]*Key Laboratory of Aerospace Thermophysics*, *Ministry of Industry and Information Technology, Harbin 150001, P. R. China*
[3]*Shandong Institute of Advanced Technology, Jinan 250100, Shandong, China*
[4]*Laboratoire Charles Coulomb (L2C), UMR 5221 CNRS-Université de Montpellier, F- 34095 Montpellier, France*
[5]*Institut Universitaire de France, 1 rue Descartes, F-75231 Paris, France*

Twisted two-dimensional bilayer materials exhibit many exotic physical phenomena. Manipulating the 'twist angle' between the two layers enables fine control of the physical structure, resulting in development of many novel physics, such as the magic-angle flat-band superconductivity, the formation of moiré exciton and interlayer magnetism. Here, combined with analogous principles, we study theoretically the near-field radiative heat transfer (NFRHT) between two twisted hyperbolic systems. This two twisted hyperbolic systems are mirror images of each other. Each twisted hyperbolic system is composed of two graphene gratings, where there is an angle $\varphi$ between this two graphene gratings. By analyzing the photonic transmission coefficient as well as the plasmon dispersion relation of twisted hyperbolic system, we prove that the topological transitions of the surface state at a special angle (from open (hyperbolic) to closed (elliptical) contours) can modulate efficiently the radiative heat transfer. Meanwhile the role of the thickness of dielectric spacer and vacuum gap on the manipulating the topological transitions of the surface state and the NFRHT are also discussed. We predict the hysteresis effect of topological transitions at a larger vacuum gap, and demonstrate that as thickness of dielectric spacer increase, the transition from the enhancement effect of heat transfer caused by the twisted hyperbolic system to a suppression. This technology could novel mechanism and control method for NFRHT, and may open a promising pathway for highly efficient thermal management, energy harvesting, and subwavelength thermal imaging.

**I. INTRODUCTION**

Since the pioneering work of Polder and van Hove [1] it is well known that radiative heat transfer (RHT) between two bodies can be significantly enhanced in the nanoscale compared with classical radiation theory [2–4]. Thanks to the tunneling of evanescent waves, either theoretically or experimentally, via the resonant coupling of surface phonon polaritons (SPhPs) [5], surface plasmon polaritons (SPPs) [6], hyperbolic polaritons [7,8], or quasi-elliptic polaritons [9], the near-field radiative heat flux can exceed the blackbody limit by several orders of magnitude. The huge radiative heat flux in the near field opens the door to various applications like thermophotovoltaic device [10-13], thermal rectification [14,15], information processing [16] and noncontact refrigeration [17]. In these applications about energy conversion and information processing, the management of near-field radiative heat transfer (NFRHT) plays an important role. Several strategies for active control of NFRHT have been proposed [18-31]. For instance, one can tune the NFRHT by external electric gating in 2D materials [18] and ferroelectric materials [19]. Or one can use drift current, i.e., nonreciprocal surface plasmon polaritons, to control NFRHT in graphene [20]. In addition, one can actively control NFRHT through an external magnetic field [21-24]. Examples such as near-field thermal Hall effect [21, 22], thermal magnetoresistance [23], and thermal modulation [24] were reported. Other schemes such as temperature control in phase-change materials [25-29], and mechanical strain [30,31] are also investigated. Despite these strategies have largely developed methods for regulating and enhancing near-field thermal radiation through the surface polaritons, a novel way for controlling NFRHT is still highly desired for near-field energy conversion and active thermal management.

---

* Corresponding author. E-mail address: yihongliang@hit.edu.cn
† Corresponding author. E-mail address: mauro.antezza@umontpellier.fr

Recently, twisted superlattice have been successfully demonstrated in bilayer two-dimensional (2D) materials, in which one layer is rotated with respect to the other, unveiling phenomena such as superconductivity at magic angle [32], topological excitons [33,34], and atomic photonic crystals [35,36]. These strong dependences on the rotation angle has opened the emerging research direction of "twistronic". Inspired by the twistronic concepts, we explore a controllable thermal transfer based on twisted hyperbolic system, which an interesting and novel modulatory mechanism may be introduced for NFRHT. The twisted hyperbolic system is composed of stacked uniaxial supporting hyperbolic polaritons. In contrast to the monolayer metasurface, due to the hyperbolic plasmon hybridization between closely adjacent metasurface, twisted hyperbolic system open new opportunities to develop a novel control method of surface polarization and NFRHT. The uniaxial hyperbolic metasurfaces can been demonstrated in several material platforms, from Waals materials [37], graphene nanoribbons [38], black phosphorus [39], and other 2D materials [40]. We consider densely packed graphene nanoribbons as the platform of choice, due to graphene nanoribbons that can provide the robust hyperbolic polaritons in a board frequency band. Other metasurface geometries may be equally viable, as a function of the wavelength range of interest.

In this work, we theoretically investigate the active control of NFRHT between two bilayer graphene nanoribbons through precise control of the polaritons dispersion through tailored interlayer coupling in twisted bilayer. Such control is dependent on the topological transitions of hyperbolic polaritons at transitional angles, that is, the polaritons change markedly in nature from hyperbolic (open) to elliptical (closed). This topological transition of polaritons is determined by the number of anti-crossing points of the dispersion lines of each isolated layer in bilayer, analogous to a Lifshitz transition in electronics. In addition, the properties of NFRHT with topological polaritons due to the thickness of dielectric spacer and vacuum gap are discussed at the end.

## II. THEORETICAL ASPECTS

Let us consider a system composed of two twisted hyperbolic systems brought into close proximity with a vacuum gap size of $d_0$ as sketched in Fig. 1(a). We can form a twisted hyperbolic system by coupling two identical graphene nanoribbons, as sketched in Fig. 1(a), separated by thickness $d_s$ of dielectric spacer. Each graphene strip has width $W$ and an air gap $G$ separates neighboring strips. We define the graphene strips of twisted hyperbolic system above the vacuum gap as $G_1$ and $G_2$ from top to bottom, and the graphene strips of twisted hyperbolic system below the vacuum gap as $G_3$ and $G_4$ from top to bottom. The upper structure is the emitter ($G_1$ and $G_2$) with a higher temperature $T_1$ and the bottom structure ($G_3$ and $G_4$) is the receiver with a lower temperature $T_2$. The temperatures are set to be $T_1 = 310$ K and $T_2 = 300$ K, respectively. Here, $\alpha$ and $\beta$ are defined perpendicular to periodical directions of the first layer and the second layer in the one bilayer, respectively. Here, the first layer represents the graphene strips close to the vacuum gap, and the second layer represents the strips away from the vacuum gap. When the second layer in one bilayer system can be rotated with respect to $y$-axis by an angle of $\varphi$ from 0° to 90°, as shown in Fig. 1(b) and (c), an interesting effects of twisted periodicities are achieved. For simplicity, we assume that the emitter and receiver are mirror images of each other and the background materials to be vacuum. On the other word, we rotate simultaneously an angle of $\varphi$ between $G_1$ and $G_2$ and an angle of $\varphi$ between $G_3$ and $G_4$. Here, we strictly compliance the emitter and receiver are mirror images of each other, that is, keeping $G_2$ parallel to $G_3$ and $G_1$ parallel to $G_4$. The non-mirrored system and dielectric backgrounds can be easily captured by generalizing the formulation in this work.

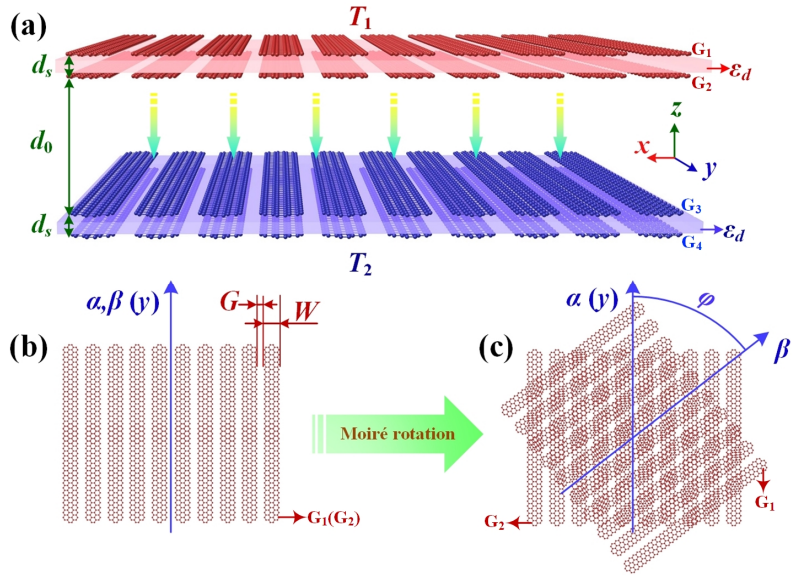

FIG. 1(a). Schematic of near-field heat transfer between two twisted hyperbolic systems. The twisted

hyperbolic system is composed of two identical graphene nanoribbons, whose width and air gap are $W$ and $G$, respectively. The structure above the vacuum gap ($G_1$ and $G_2$) is the emitter with a higher temperature $T_1$ (310 K) and the structure below the vacuum gap ($G_3$ and $G_4$) is the receiver with a lower temperature $T_2$ (300 K). (b) Top view of the crystalline structure without rotation. (c) Top view of the twisted hyperbolic system at a rotation angle $\varphi$ of with respect to $y$-axis.

The optical conductivity of the arraying graphene nanoribbons of the paper has been derived using a well-known effective medium theory (EMT) [41-43] that holds when the unit-cell period $P=W+G$ is much smaller than the operating wavelength, i.e., $P \ll \lambda_0$. Meanwhile, in references [38, 44], EMT is valid only when the vacuum gap is larger than the period of the ribbons. According to Ref [38] when the vacuum gap $d_0$ is 50 nm and the period $P$ is 10 nm, the error of EMT is less than 5% compared with Rigorous coupled-wave analysis (RCWA). As the gap distance increase, the accuracy of EMT can be further improved. Accord to EMT, this technique is based on modeling the strip near-field coupling as an effective capacitance [38]

$$\sigma_{eff} = \begin{bmatrix} \sigma_{xx} & 0 \\ 0 & \sigma_{yy} \end{bmatrix} = \begin{bmatrix} \dfrac{W\sigma_G\sigma_C}{P\sigma_C + W\sigma_G} & 0 \\ 0 & \dfrac{W}{P}\sigma_G \end{bmatrix} \quad (1)$$

where, $\sigma_C = -i\omega\varepsilon_0 P \big/ \left\{ \pi \ln\left[ \csc\left[ 0.5\pi(P-W)/P \right] \right] \right\}$ is the effective strip conductivity taking into account near-field coupling and nonlocality. In the hyperbolic regime, the metasurface acts as a metal for one transverse field polarization (Im[$\sigma_{yy}$] > 0) and as a dielectric for the other one (Im[$\sigma_{xx}$] < 0). $\omega$ is the frequency, and $\varepsilon_0$ is the vacuum permittivity. $\sigma_G$ is the optical conductivity of the graphene, and it is given by the well-known random phase approximation. Following Ref. [38], the conductivity can be written as a sum of intra-band (Drude) and inter-band contributions, i.e., $\sigma_0 = \sigma_D + \sigma_I$, respectively,

$$\sigma_D = \dfrac{i}{\omega + i/\tau} \dfrac{2e^2 k_B T}{\pi \hbar^2} \ln\left( 2\cosh\dfrac{E_f}{2k_B T} \right) \quad (2a)$$

$$\sigma_I = \dfrac{e^2}{4\hbar} \left[ G\left(\dfrac{\hbar\omega}{2}\right) + i\dfrac{4\hbar\omega}{\pi} \int_0^\infty \dfrac{G(\xi) - G(\hbar\omega/2)}{(\hbar\omega)^2 - 4\xi^2} d\xi \right] \quad (2b)$$

where $G(\xi) = \sinh(\xi/k_B T) \big/ \left[ \cosh(E_f/k_B T) + \cosh(\xi/k_B T) \right]$. The conductivity depends explicitly on the temperature $T$ and the Fermi energy $E_f$. The relaxation time $\tau$ is fixed at $10^{-13}$ s [38].

The radiative heat flux between two twisted hyperbolic systems can be calculated based on the fluctuation-dissipation theory [19]

$$q = \int_0^\infty q_\omega(\omega) d\omega = \int_0^\infty \left( \Theta(\omega, T_1) - \Theta(\omega, T_2) \right) d\omega \int_{-\infty}^\infty \int_{-\infty}^\infty \dfrac{\xi(\omega, k_x, k_y)}{8\pi^3} dk_x dk_y \quad (3)$$

where $\Theta(\omega,T) = \hbar\omega/[\exp(\hbar\omega/k_BT) - 1]$ is the mean energy of a Planck oscillator at angular frequency $\omega$. $\xi(\omega,k_x,k_y)$ is the photonic transmission coefficient (PTC) that describes the probability of photons excited by thermal energy, which can be written as

$$\xi(\omega,k_x,k_y) = \begin{cases} \mathrm{Tr}[(\mathbf{I} - \mathbf{R}_2^*\mathbf{R}_2 - \mathbf{T}_2^*\mathbf{T}_2)\mathbf{D}(\mathbf{I} - \mathbf{R}_1\mathbf{R}_1^* - \mathbf{T}_1^*\mathbf{T}_1)\mathbf{D}^*], & k < k_0 \\ \mathrm{Tr}[(\mathbf{R}_2^* - \mathbf{R}_2)\mathbf{D}(\mathbf{R}_1 - \mathbf{R}_1^*)\mathbf{D}^*]e^{-2|k_z|d}, & k > k_0 \end{cases} \quad (4)$$

for propagating ($k < k_0$) and evanescent ($k > k_0$) waves where $k = \sqrt{k_x^2 + k_y^2}$ is the surface parallel wavevector and $k_0 = \omega/c$ is the wavevector in vacuum. $k_z = \sqrt{k_0^2 - k^2}$ is the tangential wavevector along $z$ direction in vacuum and * signifies the complex conjugate. The 2×2 matrix $\mathbf{D}$ is defined as $\mathbf{D} = (\mathbf{I} - \mathbf{R}_1\mathbf{R}_2 e^{2ik_z d})$ which describes the usual Fabry-Perot-like denominator resulting from the multiple scattering between the two interfaces of receiver and emitter. The reflection matrix $\mathbf{R}$ is a 2×2 matrix in the polarization representation, which the calculation process of $\mathbf{R}$ can refer to the Appendix.

## III. HYPERBOLIC HYBRIDIZATION GUIDED HEAT TRANSFER BETWEEN TWO BILAYER GRAPHENE GRATING SYSTEMS

We first discuss the radiative heat flux (RHF) between the two bilayer graphene grating systems as a function of the rotation angle $\varphi$ by fixing the vacuum gap $d_0$ to the value of 50 nm as shown in Fig 2. Moreover, we fix the thickness of dielectric spacer to $d_s = 1$ nm. In the nature, the width of graphene stripe should be a multiple of the lattice constant $a$ of graphene ($a$=0.246 nm). We consider the width $W$ is 24 $a$ (5.9 nm) and the unit-cell period $P$ is 40$a$ (9.84 nm) to realistically depict graphene grating. Besides, we also illustrate the impact of the dielectric permittivity $\varepsilon_s$ of spacer with respect to the rotation angle $\varphi$ on the results.

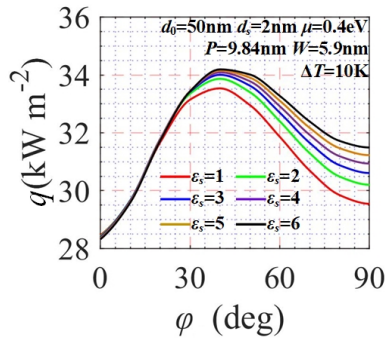

Fig. 2. Radiative heat flux $q$ as a function of rotation angle $\varphi$ at different dielectric permittivity of spacer. The parameters are $d_0$ =50 nm, $d_s$ = 1 nm, $T_1$ = 310 K, $T_2$ = 300 K, $\mu$ = 0.4 eV, $W$ =5.9 nm, and $P$ =9.84 nm.

We can see in the Fig. 2 that the presence of the rotation angle significantly modify the RHF between the two bilayer graphene grating systems. In most of the cases, the RHF is above that of in the absence of the rotation angle that means that an enhancement of heat transfer is achieved. As shown in Fig. 2, interestingly, we find that the RHF exhibits a non-monotonic dependency versus the rotation angle, especially for a lower dielectric permittivity of spacer. The maximum of RHF is observed at $\varphi \approx 40°$, which implies that the structure of twisted hyperbolic system can effectively modulate the NFRHT. For the dielectric permittivity of spacer of 1, when the rotation angle reaches 40°, the maximum of the RHF increases from 28.3 kW m$^{-2}$ for $\varphi = 0°$ to around 33.5 kW m$^{-2}$ for $\varphi = 40°$, and decreases rapidly as rotation angle further increases. In addition, it can be seen that although the dielectric permittivity of spacer between two graphene gratings could tune the RHF of meta-structure, the dependency of RHF versus the rotation angle under different dielectric permittivity has been constant. In order to conveniently reveal the physical mechanism of how the twisted hyperbolic system modulate the NFRHT, we fixed the dielectric permittivity of the spacer to 1 in the subsequent analysis.

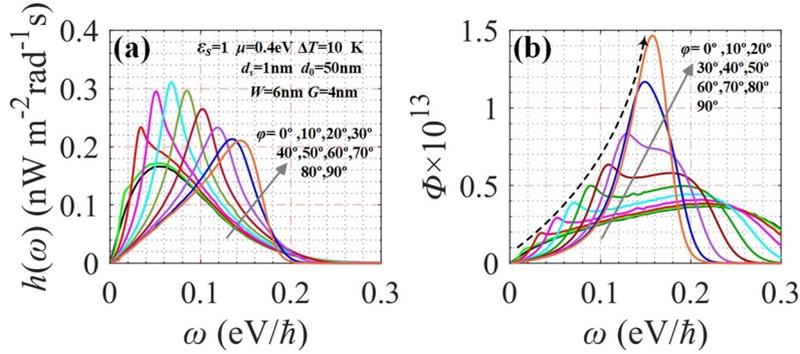

Fig. 3(a) Spectral RHF as a function of the frequency for a vacuum gap $d_0 = 50$ nm and a chemical potential $\mu = 0.4$ eV. The different lines correspond to different rotation angles. (b) Contour plots of energy transfer function $\Phi(\omega)$, as a function of frequency with different rotation angles.

To get further insight into the role of the twisted hyperbolic system, we show in Fig. 3(a) the spectral RHF for a gap $d_0=50$ nm, the dielectric permittivity of the spacer is 1, and different rotation angles. This spectral RHF is defined as the RHF per unit of frequency. It can be seen that a new peak appears as twisted hyperbolic system emerge ($\varphi > 0°$). In addition, when the rotation angle increase to 20°, the new peak begin to dominate the spectral RHF, that is, the new peak become the maximum of the spectral RHF. The new peak of the spectral RHF is blue-shifted upon increasing the rotation angle from 0.018 eV/$\hbar$ for $\varphi = 10°$ up to around 0.145 eV/$\hbar$ for $\varphi = 90°$. Notice also that the maximum of the spectral RHF also increases drastically with the rotation angle, reaching a maximum of 0.31 nW m$^{-2}$ rad$^{-1}$ s at $\varphi = 40°$. As the rotation angle further increase, the maximum of the spectral RHF gradually decrease to the value of 0.21 nW m$^{-2}$ rad$^{-1}$ s at $\varphi = 90°$.

In order to visualize the NFRHT of the rotation angle, in Fig. 3 (b), we calculate the energy transfer function $\Phi(\omega)$, given by $\Phi(\omega) = q_\omega / \left( \Theta(\omega, T_1) - \Theta(\omega, T_2) \right)$. As the rotation angle increases, the peak of energy transfer function generated by twisted hyperbolic system increases drastically, reaching a maximum of $1.5 \times 10^{13}$. This also explains that the spectral RHF in Fig. 3(b) increases as the rotation angle increases at the lower rotation angle range. However, the peak of energy transfer function generated by higher rotation angle is excited only by higher photonic energy, which corresponds to the blue-shifted peak of spectral RHF shown in the inset of Fig. 3(a). Unfortunately, the contribution of spectral RHF with higher frequency (higher photonic energy) to RHF are negligible, due to the decaying exponentially of mean energy of a Planck oscillator $\Theta(\omega,T)$ at room temperature. Therefore, in Fig. 3(a), when the rotation angle is 90°, it is difficult to observe the enhance of spectral RHF excited by the ultra-high energy transfer coefficient, which is in agreement with the declining trend of the RHF at the larger rotation angle range in Fig. 2.

## IV. UNDERLYING PHYSICS OF THE HYPERBOLIC HYBRIDIZATION

It is well known that the photonic transmission coefficient (PTC) and dispersion relation can visually reveal the physics behind effect of surface state on NFRHT. We thus employed these two physical quantities to explore the underlying physical mechanisms for this twisted hyperbolic system to the NFRHT. For a frequency of 0.15 eV/$\hbar$, the below panels of Figs. 4(b)-(f) show the distributions of the photonic transmission coefficient and the dispersion relation with rotation angle $\varphi = 0°$, 30°, 60°, 80° and 90°, respectively. As we all know, different from conventional circles, the surface state of anisotropic SPPs supported by graphite gratings graphs as a hyperbola, where the dominant contribution to the optical response of graphene comes from the scattering of free electrons. As shown in Fig. 4(b), the regime featured with PTC for bilayer graphene grating system is axisymmetric hyperbolic lines. The dispersion relation of effective graphene gratings can well reflect this phenomenon, and it is given by Eq. (A11). Moreover, as the two bodies are brought into proximity, the evanescent field of HSPPs associated with two graphene gratings can interact with each other, leading to the two dispersion curves at the higher wavevector region and the smaller wavevector region [45]. However, due to the existence of the attenuation length ($\delta_z = 1/\text{Im}(k_z)$) [19], the surface state dominated by dispersion at the higher wavevector region is easily filtered by the vacuum gap, and cannot make a corresponding contribution to the NFRHT. Therefore, in the subsequent analysis, we only show the dispersion relationship at the smaller wavevector region, unless otherwise stated.

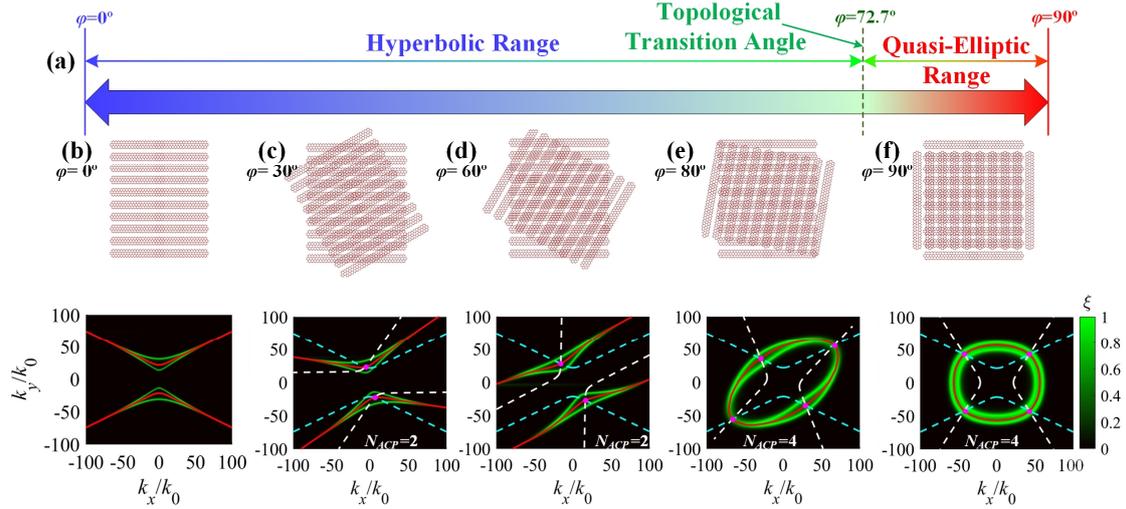

Fig. 4 Rotation-induced topological transition of plasmon polaritons. (a) Topological nature of the plasmon polaritons dispersion with respect to the rotation angle at a frequency of 0.15 eV/$\hbar$. (b)-(f) The top row shows the formed twisted fringe patterns of twisted bilayer hyperbolic system. The bottom row is the numerically simulated photonic transmission coefficient in wavevector space. The blue dashed lines and the white dashed lines are the dispersion lines of the top grating (no rotation) and bottom grating (with rotation) of twisted bilayer system. The red solid lines are the dispersion after hybridization of twisted hyperbolic system.

As shown in Figs. 4(b)-(f), it can be clearly seen that when the second individual graphene grating of bilayer system emerges mechanically rotation, the bright branch of the polaritons change markedly—from hyperbolic (open) to elliptical (closed). This photonic topological transition is governed by the coupling of in-plane hyperbolic SPPs that are individually supported by each layer of bilayer system [33,34]. Before the rotation angle at which the topological transition arises, as the rotation angle increases gradually the hyperbolic bright branch would be flattened. After a topological transition occurs on the surface state (i.e., the bright band closes), the surface state gradually change from quasi-elliptical plasmon to quasi-isotropic plasmon with the rotation angle increase. Moreover, as its surface state changes from open to closed, the topological transition yield diffraction less and low-loss directional SPPs canalization, which in turn produces a more intense surface state, i.e., a more brighter and stronger bright branch, as shown in Figs. 4(b)-(f). These results contributes to a further confirmation and explanation of the drastic increase of energy transfer function in Fig. 3(b). Furthermore, it should be emphasized that the topological transition of surface state is completely different from the previous work about controlling NFRHT based on the mechanical rotating between the emitter and the receiver. The topological transition of surface state is caused by the twisted hybridization based on the twisted hyperbolic system, which it can be seen as a nonlinear change with richer physical connotations. The previous work about the mechanical rotating between the emitter and the receiver is just a simple superposition of surface state, and the corresponding understanding is linear change [46-50].

In addition, to explore thoroughly these effects, Figs. 4(b)-(f) shows the isofrequency dispersion curves for twisted hyperbolic system considering different rotation angles. The red solid lines show the calculated dispersion bands of twisted hyperbolic system. The blue and white dashed lines correspond to the dispersion curves of the isolated top (no rotation) and bottom (with rotation) graphene gratings. This red solid line nicely locates at the bright branches, which unambiguously demonstrates that topological hybridization of HSPPs dominate the NFRHT in our system. In addition, we can predict the this transition of bright branch via a topological quantity—that is, the number of crossing points $N_{ACP}$ of the dispersion lines of each isolated graphene grating in bilayer system, analogous to a Lifshitz transition in electronic [33,34]. The bright branch of the topological polaritons change markedly in nature—from hyperbolic (open) to elliptical (closed)—when the integer number of crossing points $N_{ACP}$ changes from 2 to 4, as shown in Figs. 4(b)-(f). On the other word, at the twist angle at which the topological transition arises the number of crossing points must necessarily reach 4, as its topology changes from open to closed.

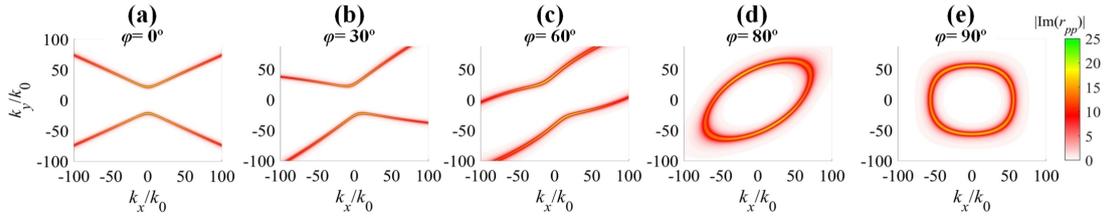

Fig. 5 Imaginary part of the reflection coefficients $r_{pp}(k_x, k_y)$ of the twisted hyperbolic system for (a) $\varphi = 0°$, (b) $\varphi = 30°$, (c) $\varphi = 60°$, (d) $\varphi = 80°$, and (e) $\varphi = 90°$ a frequency of 0.15 eV/$\hbar$. The parameters are $d_s = 1$ nm, $\mu = 0.4$ eV, $W = 5.9$ nm, $P = 9.84$ nm, and $\varepsilon_s = 1$.

Moreover, due to $p$ polarized evanescent waves occupies the main contribution in the radiation heat transfer of graphene, we calculate the reflection coefficients of the $p$ polarized evanescent waves from the vacuum to one of the twisted hyperbolic system to clearly exhibit these topological transition. Imaginary part of the reflection coefficients, is shown in Fig. 5. For the bilayer system without rotation in Fig. 5(a), the hyperbolic distribution of reflection coefficients is also presented, which agrees well with the bright branch of PTC and dispersion relation. We see that, when the rotation angle changes from 0° to 90°, the imaginary part of the reflection coefficients from hyperbolic to elliptic, experiencing a topological transition, which is both seen in the field distribution from the reflection coefficients. This means that the SPPs dispersion relations for twisted hyperbolic system are well predicted by the reflection coefficients. Meanwhile, it can be seen that as the rotation increase, this twisted hyperbolic system gradually master a stronger reflection capacity, which is both shown in the imaginary part of the reflection coefficients. These results illustrate the high tunability of NFRHT. These results also illustrate why a stronger resonance branch of PTC arise with increasing of rotation angle.

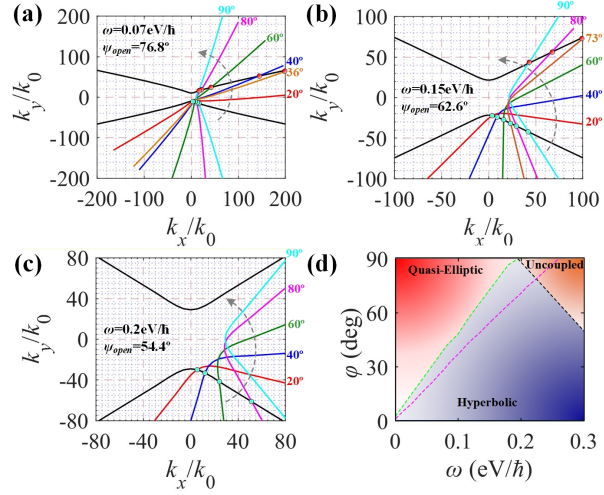

Fig. 6 Number of anti-crossing points versus the angle at the frequency of (a) 0.07 eV/ℏ, (b) 0.15 eV/ℏ, and (c) 0.2 eV/ℏ. The green circular mark indicate the location of the crossing in the fourth quadrant and the red circular mark indicate the crossing point in the first quadrant. The gray arrows show the evolution of the rotation angle. (d) Topological transition regions as a function of frequency and rotation angles. The green dashed line represents the realistic topological transition angle and the pink dashed line represents the ideal ones ($\varphi_{ideal}$ = 180°-2$\psi_{open}$).

To explain visually the topological hybridization of HSPPs and the blueshift of spectral RHF, we illustratively show the dispersion and the topological transition regions as a function of rotation angles in Fig. 6. As mentioned above, the number of crossing points as a function of the angle can powerfully predict the topological transition. Nevertheless, the number of crossing points does not show the range of transitional angles where topological transitions occur. Thus, here we utilize the the open angle $\psi_{open}$ of the hyperbolic branches of individual graphene grating of bilayer system to predict the transitional angles where topological transitions occur. For the single graphene grating of bilayer system, hyperbolic SPPs propagate at the open angle atan(Im($\sigma_{yy}$)/ Im($\sigma_{xx}$)) [51]. For the twisted hyperbolic system, the two hyperbolic bands of the individual layers hybridize and strongly couple to each other at the points in wavevector space at which they cross, leading to crossing and topological transitions. Since the hyperbolic curve is open, ideally, there is no limit on the density of states (DOS) [52]. Thus, ideally, when open angle is paired together with a given rotation angle, that is, $\varphi_{ideal}$ = 180°-2$\psi_{open}$, an anticrossing arises at the intersection of the bands, which may be easily predicted using geometrical arguments. However, practically, the hyperbolic SPPs will not hold when wavevector goes to infinity due to the existence of the attenuation length. Therefore, the angle $\varphi_{real}$ at which the actual topological transition occurs lags behind the ideal angle $\varphi_{ideal}$. For example, as shown in Fig. 6(a), we can see that for smaller rotation angles ($\varphi$ < 36°) there will only be an crossing in the fourth quadrant (green circular mark) and the complementary points in the second quadrant (not shown for simplicity). For larger angles ($\varphi >36°$), two additional crossing points emerge in the first quadrant (red circular mark) and the complementary points in the third quadrant (not shown). But, according to a frequency of 0.07 eV/$\hbar$ with an open angle of 76.8 degrees, the ideally transitional angle $\varphi_{ideal}$ only is 26.4 degrees, which the practically transitional angle $\varphi_{real}$ is much larger than the ideally transitional angle $\varphi_{ideal}$.

As the frequency increase, the open angle gradually reveal a downward trend and, so the rotation angle required for the topological transition gradually becomes larger. For example, as shown in Fig. 6(b), there only have a crossing point in the second and fourth quadrant at the larger range of rotation angle ($0<\varphi <73°$). Meanwhile, it can be clearly seen that at a higher frequency, since the hyperbolic SPPs at a large wave vector range is easily filtered by vacuum gap, the difference between the ideal transitional angles of the topological transition and the actual one is increased compared to the low frequency case, as shown in Fig. 6(d). This also led to the fact that with the further increase of frequency, which the number of crossing points can no longer up to 4 due to the finite state density in Fig. 6(c). On the other hand, as shown in Fig. 6(d), since the two hyperbolic bands of the individual layers do not couple to each other at the points in reciprocal space at which they cross, the topological transition of photonic transmission coefficient do not be generated at the higher rotation angle, in turn, forming the uncoupled surface states. This uncoupled surface state would severely inhibit NFRHT. Moreover, according to the Fig. 4, when the topological transition occurs, the brightest and strongest bright branch of surface state would be produced. As shown in Fig. 6(d), the frequency of topological transition also increases with the increasing of as the rotation angle. Thus, in Fig. 3(b), the peak of energy transfer function $\Phi(\omega)$ excited by topological transitions shows a clear blue-shift.

## IV. HYPERBOLIC HYBRIDIZATION DEPENDENCE OF THE HEAT TRANSFER FOR DIFFERENT VACUUM GAPS

In order to explore the effect of twisted hybridization on heat transfer at different vacuum gap, in Fig. 7(a), we calculate the RHF as a function of rotation angle for different vacuum gaps. The curves with different colors represent different vacuum gaps. As is well known, in the field of lower vacuum gap, the RHFs of twisted hyperbolic system in different rotation angle can be higher than the RHF at the higher vacuum gap.

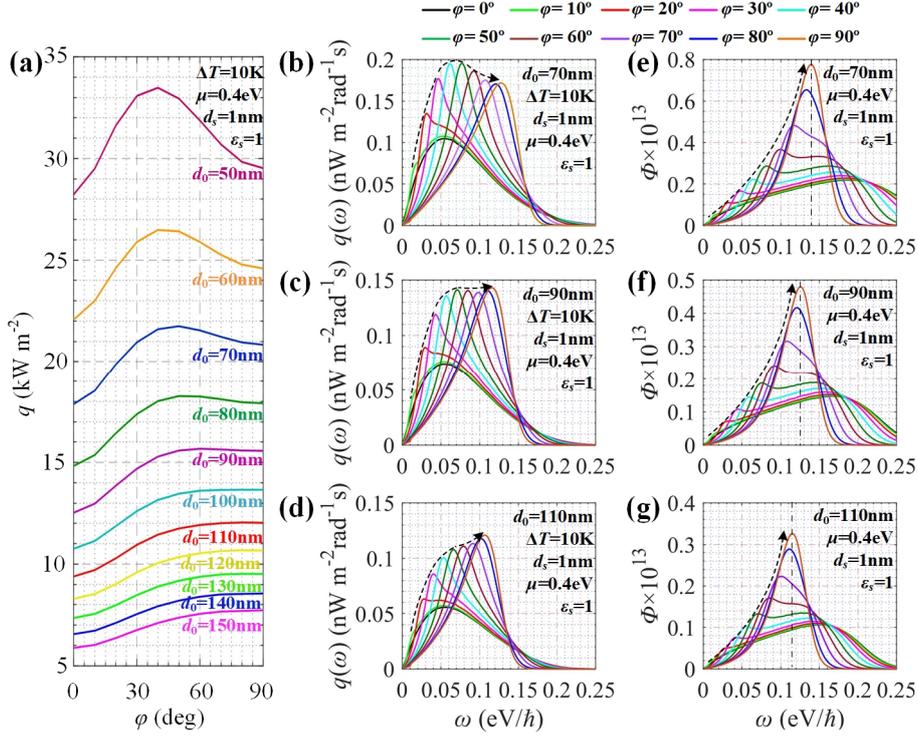

Fig. 7. (a) Radiative heat flux $q$ as a function of rotation angle $\varphi$ at different vacuum gap $d_0$. Spectral RHF as a function of the frequency with different rotation angle for a vacuum gap of (b) 70 nm, (c) 90 nm, and (d) 110 nm. Contour plots of energy transfer function $\Phi(\omega)$ as a function of frequency with different rotation angles for a vacuum gap of (e) 70 nm, (f) 90 nm, and (g) 110 nm. The parameters are $d_s = 1$ nm, $T_1 = 310$ K, $T_2 = 300$ K, $\mu = 0.4$ eV, $W = 5.9$ nm, and $P = 9.84$ nm.

We can see in the Fig. 7(a) that the presence of the rotation angle significantly modify the RHF and can enhances the RHF of system compared with the case without rotation angle, both for different vacuum gaps. In the cases of a smaller gap, i.e. vacuum gap less than 100 nm, we find that the RHF exhibits a non-monotonic dependency versus the rotation angle, especially for a lower vacuum gap. For a small vacuum gap with $d_0 = 50$ nm, we find that the non-monotonic trend is strong and the maximum of RHF is at a middle angle of $\varphi_{max} = 40°$, and the RHF decreases drastically with a further increasing of $\varphi$. However, for a higher vacuum gap, the non-monotonic trend is much weaker and the rotation angle corresponding to the maximum of RHF is larger. When the vacuum gap reach to the 100 nm, compared with the non-monotonic dependency on the rotation angle at lower vacuum gap, the larger angle can further enhance the NFRHT.

To interpret the underlying physics of the transformation of effect of twisted hybridization on heat transfer at different vacuum gap, we show in Figs. 7(b)-(d) the spectral RHF with different rotation angles. Now we show three gaps, namely $d_0$=70 nm, 90 nm and 110 nm, respectively. Physically, one can expect that the different vacuum gap would bring the different spectral response. For the vacuum gap of 70 nm, notice also that the maximum of the spectral

RHF also increases drastically with the rotation angle, reaching a maximum of 0.195 nW m$^{-2}$ rad$^{-1}$ s at $\varphi = 50°$. As the rotation angle further increase, the maximum of the spectral RHF gradually decrease to the value of 0.128 nW m$^{-2}$ rad$^{-1}$ s at $\varphi = 90°$. As the vacuum gap increases to 90 nm and the rotation angle greater than 50°, the peak of the spectral RHF is no longer modulated by the rotation angle, that is, the peak of the spectral RHF remains unchanged with increasing of rotation angle. In the plot of spectral RHF [Fig. 7 (d)], we observe that, for $d_0$=110 nm, the maximum of the spectral RHF enhances upon increasing the rotation angle, hence increasing the RHF as depicted in Fig. 7 (a). To give a further insight on the underlying physics, we show a plot of energy transfer function $\Phi(\omega)$ in Figs. 7(e)-(g). When we give a larger $d_0$ between these bilayer structures along the z-axis, a spectral redshift may be observed for the same rotation phase. Namely, as the vacuum gap increases, the frequency of topological transitions of different rotation angles would be lower. We stress that, in addition to the twisted effects inside the bilayer structure the mean energy of a Planck oscillator would play a significant role on the spectral RHF. For a lower frequency, also gradually amplify the contribution of the increase of energy transfer function generated by topological transition to the enhancement of NFRHT. The spectral redshift generated by increasing vacuum gap fortunately allows people to track intuitively the increase of energy transfer function at spectral RHF. Therefore, it can be clearly observed that with the increase of the vacuum gap, the spectral peak and the rotation angle show a positive correlation in Figs. 7(b)-(d).

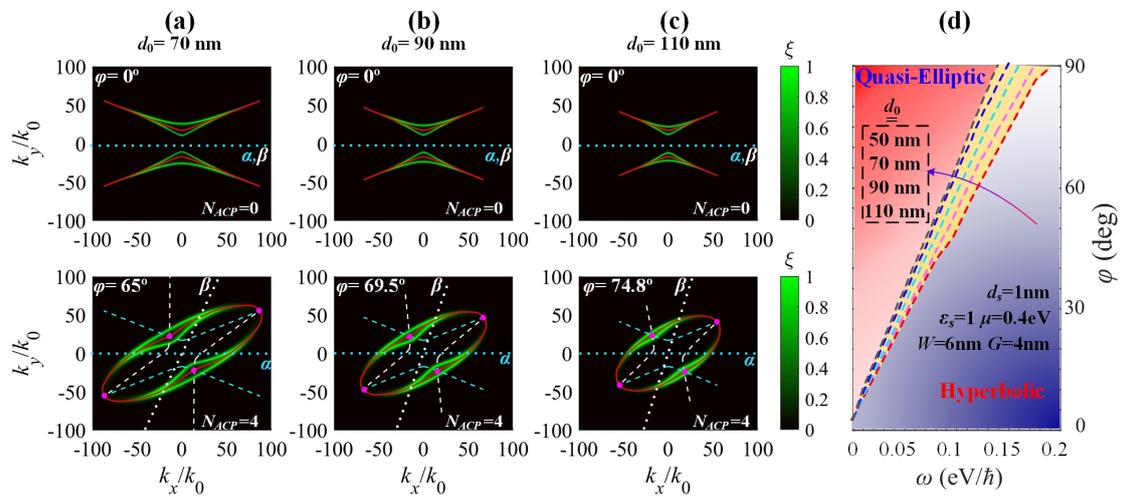

Fig. 8 (a)-(c) The top row show the numerically simulated PTCs without rotation at the vacuum gap of 70 nm, 90 nm, and 110 nm, respectively. The bottom row show the PTCs at the rotation angle of topological transitions for the vacuum gap of 70 nm, 90 nm, and 110 nm, respectively. The blue dashed lines and the white dashed lines are the dispersion lines of the top grating (no rotation) and bottom grating (with rotation) of twisted bilayer system. The red solid lines are the dispersion after hybridization of twisted hyperbolic system. The blue dotted lines and the dotted dashed lines are the optical axis of the top grating (no rotation) and bottom grating (with rotation) of twisted bilayer system. (d) Topological transition regions as a function of frequency and rotation angles at different vacuum gap.

As the vacuum gap increases, the frequency of topological transitions of rotation angles would be lower, which also means that the angle of topological transitions occur at frequencies becomes bigger. To visually reveal the physics, as shown in Fig. 8(a)-(c), we plot the PTC, dispersion relation and the corresponding rotation of the optical axis. The top row shows the PTCs without rotation for the vacuum gap of 70 nm, 90 nm, and 110 nm, respectively. It can be distinctly observed that due to the existence of the attenuation length, the wavevector range of surface state drastically shrink with increasing of vacuum gap. When the vacuum gap of system increase from 70 nm to 90 nm, the maximum wave vector of the bright branches in the positive $k_x$ quadrant significantly decrease from 90 $k_0$ to 70 $k_0$. At the vacuum gap of 110 nm, the maximum wave vector of the bright branches in the positive $k_x$ quadrant only maintain 50 $k_0$. Thus, in the bottom row of Fig. 8(a)-(c) showing the PTCs at the rotation angle of topological transitions, the surface state of HSPPs with a smaller wavevector range need to rotate a larger angle to complete the topological transition with increasing of the vacuum gap. In the Fig. 8(a)-(c), the blue and white dashed lines correspond to the dispersion curves on the bright band of surface state of the isolated top (no rotation) and bottom (with rotation) graphene gratings. Notably, as the vacuum gap increase, two dispersion curves need a lager transitional angle of topological transitions to obtain the the number of crossing points $N_{ACP}$=4, thereby achieving the change from hyperbolic to elliptical. These results also directly exhibits the delayed effect which the system at lager vacuum gap needs a lager transitional angle of topological transitions. In order to more clearly observe the transition of the surface state of the system from hyperbolic to elliptical at different vacuum gap, we plot the topological transition regions as a function of frequency and rotation angles at different vacuum gap in Fig. 8(d). The dashed lines with different colors represent the topological transition angle with different vacuum gap. In Fig. 8(d), a lager vacuum gap would pushes the topological transition angle at different frequencies to a higher range. Furthermore, in other words, a lager vacuum gap would make the topological transition frequency of different rotation angles redshifted, especially for systems with large rotation angles. These results also explain the redshift of the spectral RHF and the energy transfer function with increasing of vacuum gap in Fig. 7.

## V. HYBRIDIZATION EFFECT FOR DIFFERENT THICKNESSES OF DIELECTRIC SPACER

Note that one important aspect of the twisted hybridization is the dependence of the thickness of dielectric spacer between adjacent gratings in bilayer. As shown in Eq. A9, the interbedded coupling strength at bilayer is essentially dependent on the distance between two hyperbolic polaritons. More generally, the term $e^{-ik_{z0}d_s}$ determines the strength of the coupling. Thus, we can tune the twisted hybridization of hyperbolic polaritons via the thickness between adjacent sheets in bilayer.

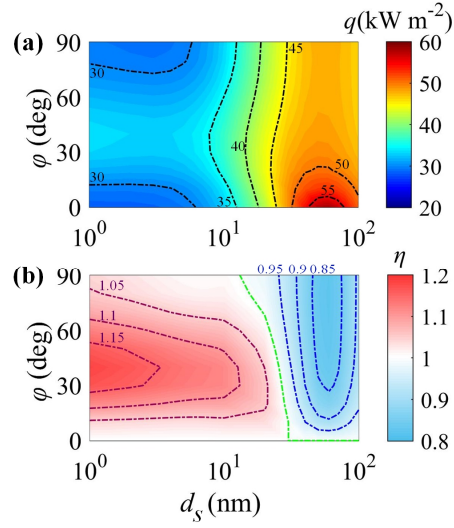

Fig. 9(a) Radiative heat flux $q$ as a function of separated thickness $d_s$ at different rotation angle $\varphi$. (b) Heat transfer coefficient ratio $\eta$ between hyperbolic system and non-rotating system as a function of separated thickness and rotation angle. The parameters are $d_0 = 50$ nm, $T_1 = 310$ K, $T_2 = 300$ K, $\mu = 0.4$ eV, $W = 5.9$ nm, and $P = 9.84$ nm.

To get insight into the role of the thickness of dielectric spacer on the NFRHT of the twisted hyperbolic system we show the radiative heat flux $q$ as a function of the thickness at $d_0 = 50$ nm for different rotation angle $\varphi$ in Fig. 9. As shown in Fig. 9(a), we find that, the RHF without rotation exhibits a s non-monotonic dependency versus the separated thickness. In fact, we find that the very compact bilayer structure reveal a suppressed heat transfer compared with the one with larger thickness. A local maximum of heat transfer is observed as the $d_0$ is comparable but a little bit larger than $d_s$. After the local maximum point, $q$ decreases, because that, as the thickness increases, the evanescent wave from the second graphene grating is gradually filtered by the vacuum gap. In order to give a more intuitive feeling of the effect on RHF from the twisted hyperbolic system with different separated thickness, Fig. 9(b) presents the RHF ratio $\eta$ between twisted hyperbolic system and non-rotating system as a function of separated thickness and rotation angle at $d_0 = 50$ nm. We see that as $d_s \leqq 10$ nm, the ratio first increases and then decreases with increasing of rotation angle. When the separated thickness is 1 nm, the maximum heat transfer coefficient ratio $\eta$ can be close to 1.2. The strongest suppression occurs when the thickness is 60 nm, which the maximum suppression can be close to 0.8.

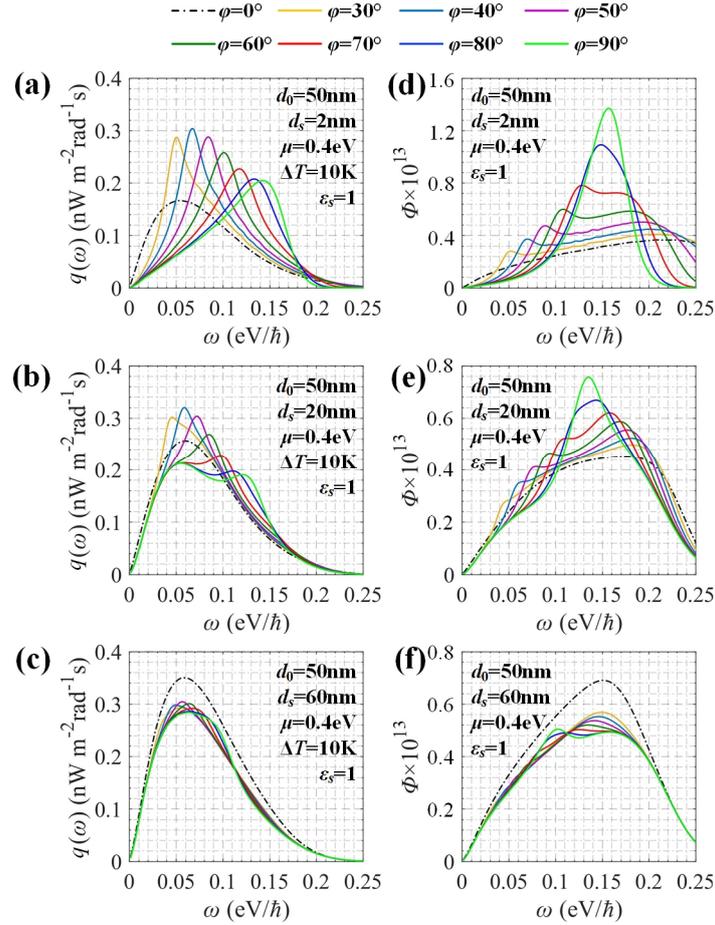

Fig. 10. Spectral RHF as a function of the frequency with different rotation angle for the thickness of dielectric spacer of (a) 2 nm, (b) 20 nm, and (c) 60 nm. Contour plots of energy transfer function $\Phi(\omega)$ as a function of frequency with different rotation angles for a vacuum gap for the thickness of dielectric spacer of (d) 2 nm, (e) 20 nm, and (f) 60 nm. The parameters are $d_0$ = 50 nm, $T_1$ = 310 K, $T_2$ = 300 K, $\mu$ = 0.4 eV, $W$ =5.9 nm, and $P$ =9.84 nm.

Now, we turn to the near-field spectral radiative heat flux. Figures 10(a)-(c) show the spectra of radiative heat flux under different rotation angle for the thickness of 2 nm, 20 nm, and 60 nm, respectively. As shown in Figs. 10(a)-(c), in the absence of rotation, we note that the contribution from the thickness of dielectric spacer to amplify spectral RHF is gradually stronger. When the the thickness of dielectric spacer reaches 2 nm from 60 nm, the maxima of the spectral RHF is increase from 0.17 nW·m$^{-2}$rad$^{-1}$·s to around 0.35 nW·m$^{-2}$rad$^{-1}$·s. However, this obvious enhancement effect on spectral RHF can be ignored at a smaller rotation angle ($\varphi$<50°) in Figs. 10(b) and (c). This also leads to for the system with a larger thickness of dielectric spacer, the increase in the RHF supported by rotation is gradually reduced, ultimately producing a suppression effect in Fig. 9. Compared with the system with smaller rotation angle, a larger thickness of dielectric spacer would lead to the obvious redshift and increasing of the spectral RHF after the rotation angle exceeds 50°. To get insight into the role of the rotation angle on the spectra of radiative heat flux with different thickness of dielectric spacer, we calculate the energy transfer function in Figs. 10(d)-(f). As a thickness of dielectric spacer with $d_s$ = 20 nm is used, we see in Fig. 10(e) that the energy transfer function gets broader but the maximum decreases compared with the ones at $d_s$ = 2 nm, especially in $\varphi$=90°. This trend is particularly obvious when the thickness of dielectric spacer increases to 60 nm. Meanwhile, it can be clearly seen that the influence of the thickness of the dielectric spacer on the energy transfer coefficient at small rotation angles ($\varphi$<50°) and low frequencies is almost negligible. As a result, effect for the thickness of the dielectric spacer on spectral RHF is very limited at a smaller rotation angle ($\varphi$<50°) in Figs. 10(b) and (c). For larger rotation angle ($\varphi$>50°), the redshift caused by the the increase in thickness result in the system to occupy a higher energy transfer coefficient at low frequencies, which in turn leads to the spectral RHF redshift and increases in Figs. 10(b) and (c). Meanwhile, as the thickness of dielectric spacer increase, the enhancement effect of twisted hyperbolic system on the energy transfer function gradually decreases. As shown in Fig. 10(b), when the rotation angle of the system at $d_s$ = 2 nm is increased from 0° to 90°, the energy transfer function at a frequency of 0.13 eV/$\hbar$ significantly increases from 0.31×10$^{13}$ to 0.94×10$^{13}$. However, at the thickness of 20 nm, the energy transfer function at a frequency of 0.13 eV/$\hbar$ increase from 0.44×10$^{13}$ to 0.75×10$^{13}$. Moreover, for the thickness of 60 nm, the energy transfer function at a frequency of 0.13 eV/$\hbar$ begins to attenuate from 0.65×10$^{13}$ to 0.48×10$^{13}$.

The origin of the dependency of the the thickness of dielectric spacer on HSPPs supported by twisted hyperbolic system can be understood with a concrete analysis of the energy transmission coefficient and the dispersion relations. In particular, we show PTC with a thickness of dielectric spacer of 2 nm, 20 nm and 60 nm, respectively, in Fig. 11. A frequency of 0.13 eV/$\hbar$ and $\varphi$=0° is considered for the left figures, and a frequency of 0.13 eV/$\hbar$ and $\varphi$=0° for the right figures. Meanwhile, we plot the plasmon dispersion relations in Fig. 11 where the white curves represent the plasmon dispersions. When the two bodies are brought into proximity, the evanescent field of HSPPs associated with two graphene gratings can interact with each other, leading to the two dispersion curves at the higher wavevector region (symmetrical dispersion) and the smaller wavevector region (anti-symmetrical dispersion) [19,38]. For $\varphi$=0°, as the dielectric spacer becomes thicker, the HSPPs bright band supported by symmetrical dispersion gradually shifts to the low wavevector area, resulting in a gradual decrease in the dissipation of the vacuum gap. As a result, the SPPs bright band supported by symmetrical dispersion gradually brighter its contribution to energy transfer gradually increases. Consequently, in the absence of rotation angle, we obtain a large RHF for a thicker dielectric spacer, as shown in Fig. 9 and Fig. 10. We can see that, for $\varphi$=90° with the increasing of thickness of dielectric spacer, the shape of the SPPs bright band and dispersion relations begins to change significantly, that is, it changes from an isotropic circle to an anisotropic rhombus. These has never been noted in the noncontact heat exchanges at nanoscale before. Meanwhile, for $\varphi=90°$, due to the coupling strength will decay when the thickness increases, the strength of SPPs bright band supported by second layer of graphene grating at bilayer system becomes weaker, which in turn leads to the enhancement effect of twisted hyperbolic system on the energy transfer function gradually decreases in Fig. 10. This is because for large separation, two graphene gratings would be very weakly coupled and the anti-crossing feature will resultantly be very weak and nearly disappear. It can be observed that the surface state degrade into the hyperbolic bright band due to the surface state supported by the edge of the rhombus dispersion perpendicular to the $k_x$-axis is already very weak. Thus, we can tune the dispersion and the SPPs of twisted hyperbolic system via the thickness of dielectric spacer.

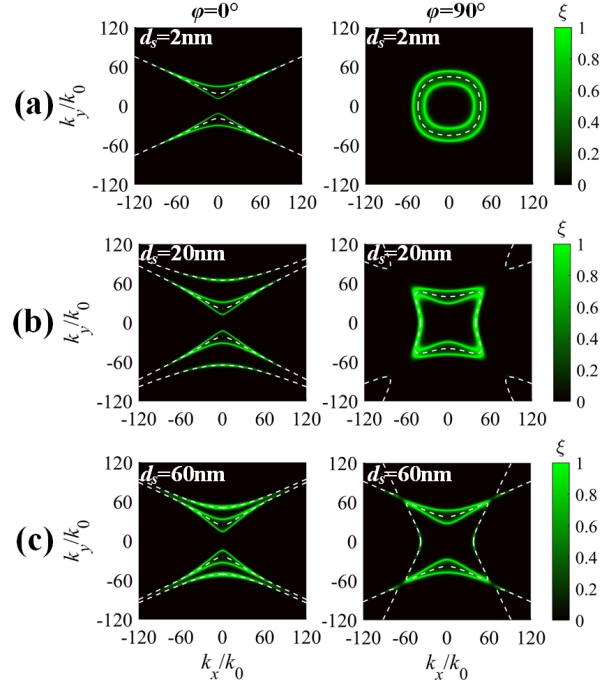

Fig. 11 (a)-(c) The left row show the numerically simulated PTCs without rotation at the thickness of dielectric spacer of 2 nm, 20 nm, and 60 nm, respectively. The right row show the PTCs at the rotation angle of 90° for the thickness of dielectric spacer of 2 nm, 20 nm, and 60 nm, respectively. The white dashed lines are the dispersion after hybridization of hyperbolic system.

To show thoroughly these effects, Figure 12 shows the isofrequency dispersion curves for superlattice considering the thickness of dielectric spacer at the frequency of 0.13 eV/$\hbar$. In the thickness of dielectric spacer at $d_s$ = 2 nm, since all supported SPPs exhibit identical characteristics independently of their propagation direction, SPPs is isotropic and thus possesses a circular IFC. As the thickness of dielectric spacer increases, due to two evanescently coupled polaritons with respect to one another in the bilayer system, the shape of the IFC gradually changes from circular to quasi-rhombus. With a thicker dielectric spacer, the superlattice supports higher $k$ surface states. Meanwhile, as seen in in Fig 12, the hyperbolic branches in the corners is visible and move towards the low wavevector area with the increasing thickness of dielectric spacer.

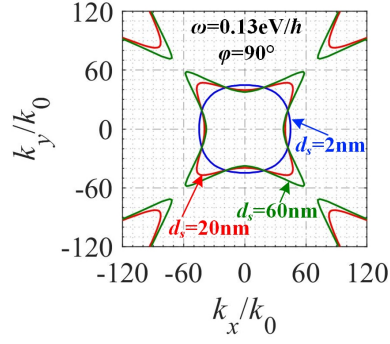

Fig. 12 Dispersion dependence on the thickness of dielectric spacer at the frequency of 0.13 eV/$\hbar$ Thickness essentially determines the coupling strength. Here, the mutual angle is 90°.

## VI. CONCLUSIONS

In summary, topological transition of plasmon polaritons in near-field radiative heat transfer between twisted hyperbolic systems composed of bilayer of twisted graphene gratings is revealed. The HSPPs topology of the twisted hyperbolic system we have used could be tuned from open (hyperbolic) to closed (elliptical) contours at a photonic transitional angle, which in turn modulates effectively the radiative heat transfer. The underlying physics are interpreted qualitatively through analyzing the dispersion of the individual metasurfaces in the wave-vector space as well as peculiar anticrossing features linked by the topological nature of the number of anti-crossing points, clearly highlighting the role played by the rotation angle. Moreover, we have found that due to the dissipative characteristics of evanescent waves in vacuum, the transitional angle of topological transition lags behind the angle predicted by the open angle of the independent hyperbolic branches. This hysteresis effect of topological transitions would be more obvious with increasing of vacuum gap. Finally, we find the transition from the enhancement effect of heat transfer caused by the twisted hyperbolic system to a suppression one with increasing of thickness of dielectric spacer. Interestingly, the thickness of dielectric spacer would distinctly modify the shape of topological polaritons. For example, at a rotation angle of 90°, there exists a modes evolution of topological polaritons with different thickness of dielectric space (from circular to quasi-rhombus) due to the change of the anti-crossing feature, which has never been noted in the noncontact heat exchanges at nanoscale before.

Our work represents a first step in the study of the modification of energy exchanges mediated by twisted physics and is expected to provide a more powerful way to regulate the energy transport, meanwhile in turn, opens up a way to explore highly efficient thermal management, energy harvesting, and subwavelength thermal imaging.

## ACKNOWLEDGMENT

This work was supported by the National Natural Science Foundation of China (Grant No. 51706053) and by the Fundamental Research Funds for the Central Universities (Grant No.

AUGA5710094020).

**APPENDIX: THE REFLECTION MATRIX R OF IN MULTILAYER ANISOTROPIC METASURFACES**

In the section, a generalized 4 × 4 T-matrix formalism for arbitrary anisotropic 2D layers is developed, from which the general relations for the surface waves dispersions and reflection matrix **R** are deduced. Let us first consider a single anisotropic metasurface at the interface between two semi-infinite media. Within the homogenization approach, the EM response of such a metasurface, in general, can be described by a fully populated conductivity tensor $\hat{\sigma}''$ in the wave-vector space. Besides, $\hat{\sigma}''$ denotes the conductivity tensor in the wave-vector space [29],

$$\hat{\sigma}'' = \begin{pmatrix} \sigma''_{xx} & \sigma''_{xy} \\ \sigma''_{yx} & \sigma''_{yy} \end{pmatrix} = \frac{1}{k^2}\begin{pmatrix} k_x^2\sigma_{xx} + k_y^2\sigma_{yy} + k_xk_y(\sigma_{xy}+\sigma_{yx}) & k_x^2\sigma_{xy} - k_y^2\sigma_{yx} + k_xk_y(\sigma_{yy}-\sigma_{xx}) \\ k_x^2\sigma_{yx} - k_y^2\sigma_{xy} + k_xk_y(\sigma_{yy}-\sigma_{xx}) & k_x^2\sigma_{yy} + k_y^2\sigma_{xx} - k_xk_y(\sigma_{xy}+\sigma_{yx}) \end{pmatrix} \quad (A1)$$

Following Refs. [53-55] let us write separately the EM field of the *p*-polarized (TM) and *s*-polarized (TE) components of the EM wave, which then will be mixed by the non-diagonal response of a metasurface. The *p* waves with the magnetic field perpendicular to the plane of incidence possess the EM field components $E_p = \{E_x,0,E_z\}$, $H_p = \{0,H_y,0\}$. For the *s* waves, with the electric field perpendicular to the plane of incidence, the EM field components are $E_s = \{0,E_y,0\}$, $H_s = \{H_x,0,H_z\}$. Substituting *p*-wave and *s*-wave into boundary conditions of metasurface [56], one obtain the 4 × 4 T-matrix, which gives the relation between all the electric field components in the media above and below the metasurface:

$$\begin{bmatrix} H_{p1}^+ \\ H_{p1}^- \\ E_{s1}^+ \\ E_{s1}^- \end{bmatrix} = \hat{T}_{1\to 2} \begin{bmatrix} H_{p2}^+ \\ H_{p2}^- \\ E_{s2}^+ \\ E_{s2}^- \end{bmatrix} \quad (A2)$$

here, the sign of + and - represent forward and backward wave. $\hat{T}_{1\to 2}$ can be derived as:

$$\hat{T}_{1\to 2} = \frac{1}{2}\begin{bmatrix} \frac{k_{z2}}{\varepsilon_2}\begin{bmatrix} P^{++} & P^{--} \\ P^{-+} & P^{+-} \end{bmatrix} & \sqrt{\frac{\mu_0}{\varepsilon_0}}\sigma''_{xy}\begin{bmatrix} 1 & 1 \\ 1 & 1 \end{bmatrix} \\ \sqrt{\frac{\mu_0}{\varepsilon_0}}\frac{\mu_1 k_{z2}}{\varepsilon_2 k_{z1}}\sigma''_{yx}\begin{bmatrix} 1 & -1 \\ -1 & 1 \end{bmatrix} & \frac{\mu_1}{k_{z1}}\begin{bmatrix} S^{++} & S^{-+} \\ S^{--} & S^{+-} \end{bmatrix} \end{bmatrix} \quad (A3)$$

where the *p*-waves components are

$$P^{++} = \frac{\varepsilon_1}{k_{z1}} + \frac{\varepsilon_2}{k_{z2}} + \frac{\sigma''_{xx}}{\omega\varepsilon_0} \;;\; P^{--} = -\frac{\varepsilon_1}{k_{z1}} + \frac{\varepsilon_2}{k_{z2}} - \frac{\sigma''_{xx}}{\omega\varepsilon_0}$$

$$P^{-+} = -\frac{\varepsilon_1}{k_{z1}} + \frac{\varepsilon_2}{k_{z2}} + \frac{\sigma''_{xx}}{\omega\varepsilon_0};\; P^{+-} = \frac{\varepsilon_1}{k_{z1}} + \frac{\varepsilon_2}{k_{z2}} - \frac{\sigma''_{xx}}{\omega\varepsilon_0} \quad (A4)$$

where the *s*-waves components are

$$S^{++} = \frac{k_{z1}}{\mu_1} + \frac{k_{z2}}{\mu_2} + \omega\mu_0\sigma''_{yy}; \quad S^{-+} = \frac{k_{z1}}{\mu_1} - \frac{k_{z2}}{\mu_2} + \omega\mu_0\sigma''_{yy}$$
$$S^{--} = \frac{k_{z1}}{\mu_1} - \frac{k_{z2}}{\mu_2} - \omega\mu_0\sigma''_{yy}; \quad S^{+-} = \frac{k_{z1}}{\mu_1} + \frac{k_{z2}}{\mu_2} - \omega\mu_0\sigma''_{yy} \tag{A5}$$

here, subscripts 1 and 2 represent the media above and below the metasurface, respectively. $k_z$, $\mu$, and $\varepsilon$ are tangential wavevector along $z$ direction in media, magnetic permeability of media, and dielectric function of media. Subscripts $p$ and $s$ represent the $p$-polarized (TM) and $s$-polarized (TE), respectively. $\varepsilon_0$ is the vacuum permittivity; $\mu_0$ is the vacuum permeability. In general, for any 4 × 4 T-matrix that links all the electric field components in a first layer with those in $N$ layer,

$$\begin{bmatrix} H^+_{p1} \\ H^-_{p1} \\ E^+_{s1} \\ E^-_{s1} \end{bmatrix} = \begin{bmatrix} T_{11} & T_{12} & T_{13} & T_{14} \\ T_{21} & T_{22} & T_{23} & T_{24} \\ T_{31} & T_{32} & T_{33} & T_{34} \\ T_{41} & T_{42} & T_{43} & T_{44} \end{bmatrix} \begin{bmatrix} H^+_{p2} \\ H^-_{p2} \\ E^+_{s2} \\ E^-_{s2} \end{bmatrix} \tag{A6}$$

The reflection matrix **R** is defined and expressed in terms of the T-matrix elements as follows [57]:

$$r_{pp} = \frac{H^-_{p1}}{H^+_{p1}}\bigg|_{E^+_{s1}=0} = \frac{T_{21}T_{33}-T_{23}T_{31}}{T_{11}T_{33}-T_{13}T_{31}}, \quad r_{ps} = \frac{E^-_{s1}}{H^+_{p1}}\bigg|_{E^+_{s1}=0} = \frac{T_{41}T_{33}-T_{43}T_{31}}{T_{11}T_{33}-T_{13}T_{31}}$$
$$r_{sp} = \frac{H^-_{p1}}{E^+_{s1}}\bigg|_{H^+_{p1}=0} = \frac{T_{11}T_{23}-T_{13}T_{21}}{T_{11}T_{33}-T_{13}T_{31}}, \quad r_{ss} = \frac{E^-_{s1}}{E^+_{s1}}\bigg|_{H^+_{p1}=0} = \frac{T_{11}T_{43}-T_{13}T_{41}}{T_{11}T_{33}-T_{13}T_{31}} \tag{A7}$$

The formalism developed above can be easily generalized for an arbitrary number of layers by multiplying the T-matrices corresponding to each layer. For a multilayer metasurface consisting of $N$ 2D-layers with effective conductivity tensors $\hat{\sigma}^{"}_j$ ($i = 1,2,...,N$). One obtain the 4 × 4 T-matrix, which gives the relations between all the electric field components and the magnetic field components in the multilayer metasurface:

$$\begin{bmatrix} H^+_{p1} \\ H^-_{p1} \\ E^+_{s1} \\ E^-_{s1} \end{bmatrix} = \hat{T}_{1\to N} \begin{bmatrix} H^+_{pN} \\ H^-_{pN} \\ E^+_{sN} \\ E^-_{sN} \end{bmatrix} \tag{A8}$$

here

$$\hat{T}_{1\to N} = \hat{T}_1 \hat{T}_{d_{i,1}} \hat{T}_2 \hat{T}_{d_{i,2}} ... \hat{T}_{d_{i,N-1}} \hat{T}_N \tag{A9}$$

where $\hat{T}_j$ is obtained from $\hat{T}_1$ [Eq. (A3)] by replacing $\hat{\sigma}^{"}_1$ with $\hat{\sigma}^{"}_j$, $\hat{T}_{d_{i,j}}$ is the T-matrices for a light propagating through the vacuum between two adjacent 2D layers with

corresponding thicknesses $d_{i,j}$ ($j = 1,2,...,N−1$):

$$T_{d_{i,j}} = \begin{bmatrix} e^{-ik_{zj}d_{i,j}} & 0 & 0 & 0 \\ 0 & e^{ik_{zj}d_{i,j}} & 0 & 0 \\ 0 & 0 & e^{-ik_{zj}d_{i,j}} & 0 \\ 0 & 0 & 0 & e^{ik_{zj}d_{i,j}} \end{bmatrix} \quad (A10)$$

The dispersion of the collective surface waves in such an *N*-layer system can be found as zeros of the denominator of the reflection coefficients:

$$T_{11}T_{33} - T_{13}T_{31} = 0 \quad (A11)$$

T-matrix allows us to obtain all necessary characteristics (the reflection and the dispersion relation) using general Eqs. (A3)–(A11).